\documentclass[letterpaper,twocolumn,english,aps, pre, showpacs, superscriptaddress]{revtex4}
\usepackage[T1]{fontenc}
\usepackage[latin1]{inputenc}
\usepackage{graphicx}

\makeatletter

\providecommand{\tabularnewline}{\\}

\usepackage{color}

\usepackage{babel}
\makeatother
\begin{document}

\title{Analysis of a microscopic stochastic model of microtubule dynamic
instability}

\author{Gennady Margolin}

\affiliation{Department of Mathematics}

\affiliation{Interdisciplinary Center for the Study of Biocomplexity, University
of Notre Dame, Notre Dame, IN 46556, USA}

\author{Ivan V. Gregoretti}

\affiliation{Department of Chemistry and Biochemistry}

\affiliation{Interdisciplinary Center for the Study of Biocomplexity, University
of Notre Dame, Notre Dame, IN 46556, USA}

\author{Holly V. Goodson}

\affiliation{Department of Chemistry and Biochemistry}

\affiliation{Interdisciplinary Center for the Study of Biocomplexity, University
of Notre Dame, Notre Dame, IN 46556, USA}

\author{Mark S. Alber}

\email{malber@nd.edu}

\affiliation{Department of Mathematics}

\affiliation{Interdisciplinary Center for the Study of Biocomplexity, University
of Notre Dame, Notre Dame, IN 46556, USA}

\begin{abstract}
A novel theoretical model of dynamic instability of a system of linear
(1D) microtubules (MTs) in a bounded domain is introduced for studying
the role of a cell edge in vivo and analyzing the effect of competition
for a limited amount of tubulin. The model differs from earlier models
in that the evolution of MTs is based on the rates of single unit
(e.g., a heterodimer per protofilament) transformations, in contrast
to postulating effective rates/frequencies of larger-scale changes,
extracted, e.g., from the length history plots of MTs. Spontaneous
GTP hydrolysis with finite rate after polymerization is assumed, and
theoretical estimates of an effective catastrophe frequency as well
as other parameters characterizing MT length distributions and cap
size are derived. We implement a simple cap model which does not include
vectorial hydrolysis. We demonstrate that our theoretical predictions,
such as steady state concentration of free tubulin, and parameters
of MT length distributions, are in agreement with the numerical simulations.
The present model establishes a quantitative link between microscopic
parameters governing the dynamics of MTs and macroscopic characteristics
of MTs in a closed system. Lastly, we use a computational Monte Carlo
model to provide an explanation for non-exponential MT length distributions
observed in experiments. In particular, we show that appearance of
such non-exponential distributions in the experiments can occur because
the true steady state has not been reached, and/or due to the presence
of a cell edge. 
\end{abstract}


\maketitle

\section{Introduction}

Microtubules (MTs) are intracellular polymers which provide a part
of the cytoskeleton and are responsible for many cell functions including
division, organelle movement, and intracellular transport. A cell
is a living object, and as such it has to constantly adjust to and
communicate with a changing environment. For this purpose, MTs possess
a property called dynamic instability, which enables them to promptly
switch between two modes, growth and shortening \cite{Mitchison84,Hill84a,Howard03}.
This is achieved through MT having a stabilizing cap which keeps the
MT from disassembling. The MT tends to depolymerize when the cap is
lost \cite{Carlier81,Hill84a,Hill84b,Howard03}. The cap gradually
hydrolyzes and becomes unstable as well, and so for the MT to survive
it has to grow to renew its cap.

The existence of a GTP cap at the end of MTs \cite{Carlier81} and
the phenomenon of dynamic instability \cite{Mitchison84} were discovered
in the early 1980s. Hill and Chen used a Monte Carlo approach to simulate
this behavior \cite{Hill84a,Chen85a}, employing a representation
of an MT in which its cap could consist of many units (heterodimers).
One of the main outcomes of their work was a suggestion that a two-phase
(cap, no cap) model of dynamic instability, based only on observable
macroscopic rates of phase and length changes, was sufficient to understand
the behavior of the ensemble of MTs (cf. \cite[figs. 4-6]{Hill84b}).
This approach has been prevalent since then in modeling the behavior
of an ensemble of MTs \cite{Dogterom93,Bolterauer99a,Bolterauer99b,Gliksman93,Govindan04,Bayley89,Hill84b,Chen85b,Freed02,Maly02}.
One exception was the theoretical model of Flyvbjerg et al. \cite{Flyvbjerg94,Flyvbjerg96}
which considered an elegant theoretical model of the GTP cap dynamics.
This model was based on microscopic constitutuve processes of spontaneous
and vectorial hydrolyses inside the MT, and fluctuating growth of
the cap size. Quite recently many detailed models of a single MT began
to emerge \cite{Janosi02,VanBuren02,VanBuren05,Stukalin04,Molodtsov05}
which try to incorporate biological details observed recently due
to advances in the experimental techniques. In particular, it is now
known from the experiments that the tips of MTs can have geometrical
configurations typical to growing and shortening MTs, which differ
from one another (e.g., \cite{Howard03}). This is closely related
to the idea of the \emph{structural}, and not necessarily a GTP cap
\cite{Janosi02}, when due to tensile stresses inside the elastic
body of an MT, its shape deforms from a cylinder towards the tip.

In both \emph{in vitro} and \emph{in vivo} experiments, dynamics of
MTs have been observed under a large variety of physical conditions
and in various chemical environments. Much data has been accumulated
including parameter values describing the MT dynamics and length distributions.
Can these values be predicted based on the conditions of the experiments?
How would change in ambient conditions or the presence of spatial
constraints affect observables? These questions are difficult, if
not impossible, to answer using the models with postulated observable
(macroscopic) rates.

In this paper we analyze a novel model of MT dynamics in a finite
domain bounded by the cell edge, which involves competition of individual
MTs for tubulin. The model is based on a linear 1D approximation of
a MT structure. We consider the role of the boundary and extend the
model to incorporate finite hydrolysis. Our model is different from
earlier works \cite{Govindan04,Maly02} addressing the role of the
edge in that we explicitly consider the concentration dependence of
the dynamic instability parameters, as well as a competition for a
limited tubulin pool.

Namely, we use a generalization of a microscopic model of MTs introduced
in \cite{Gregoretti}. Instead of postulating macroscopic rates \cite{Hill84a,Hill84b}
or deducing them from numerical simulations \cite{Chen85a}, we estimate
them theoretically from basic microscopic rates of (de)polymerization
and hydrolysis of a single unit (which can, but does not have to be
interpreted as a heterodimer). This results in a higher-resolution
theoretical model which may be more suitable for today's higher-resolution
experiments, and can partially address the above questions.

One of the main results of the paper is an establishment of a link,
by using analytical formulas, between microscopic parameters, describing
polymerization/depolymerization and hydrolysis of individual units,
and macroscopic (observable) characteristics of the MT dynamics and
ensemble. We demonstrate how to approximate macroscopic steady state
behavior of MTs using microscopic rates and vice versa, extract microscopic
rates from macroscopic behavior. Hence, it makes it possible to analytically
and quantitatively predict the effect of changes in microscopic parameters
on observable features, as well as to deduce microscopic changes from
observed changes in macroscopic behavior, when relevant geometry and
chemistry is taken into account.

We also use a computational Monte Carlo model to provide an explanation
for non-exponential MT length distributions observed in experiments.
In particular, we show that appearance of such non-exponential distributions
in the experiments can occur because the true steady state has not
been reached, and/or due to the presence of a cell edge.

\begin{figure}
\begin{center}\includegraphics[%
  width=0.4\paperwidth,
  keepaspectratio]{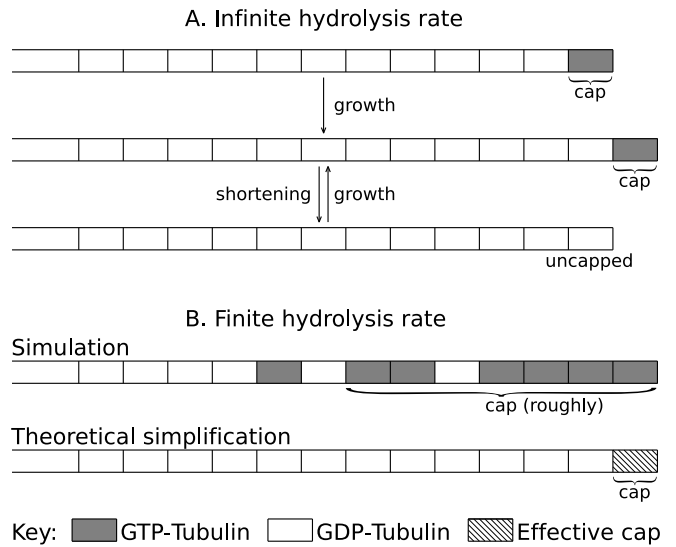}\end{center}

\caption{\label{cap:cartoon} Schematic representation of the model.}
\end{figure}

The paper is organized as follows. The conceptual model and its computational
implementation are presented in Section \ref{sec:Model-description}.
Next, in Section \ref{sec:Cap-model} we develop a cap model where
the cap can have any number of (sub)units, which can be single heterodimers
- see Figure \ref{cap:cartoon}B. This cap model differs from previous
models in that it does not involve vectorial or induced hydrolysis.
Using this model we derive approximate expressions for observable
rates. We then describe in Sections \ref{sec:Ensemble-dynamics} and
\ref{sec:Competition-for-tubulin} a quantitative theoretical analysis
of a lower-resolution model with the cap being treated as a single
unit - see Figure%
\ref{cap:cartoon}A. The influence of the cell edge is also studied
there. Section \ref{sec:Competition-for-tubulin} describes balance
between polymerized and free tubulin in a bounded domain with a fixed
total amount of tubulin present. Finally, we discuss and summarize
our findings in Sections \ref{sec:Discussion} and \ref{sec:Conclusions}.

\section{Model description, parameters and notations \label{sec:Model-description}}

In this section we describe a basic model of dynamics of MTs. We consider
a domain of size $L_{x}\times L_{y}\times L_{z}$ with $N_{n}$ available
nucleation sites for MTs in its center. $N_{n}$ is the maximal number
of MTs - cf. Table%
\begin{table*}
\begin{center}\begin{tabular}{|c|c|c|}
\hline 
Symbol&
 Definition&
 Dimensions\tabularnewline
\hline
{\small $c$}&
 {\small concentration of free tubulin}&
 {\small $\mu M$}\tabularnewline
\hline
{\small $c_{eq}^{\infty}$}&
{\small critical concentration of free tubulin }&
 {\small $\mu M$}\tabularnewline
\hline
{\small $c_{tot}$}&
 {\small total concentration of tubulin}&
 {\small $\mu M$}\tabularnewline
\hline
{\small $k_{gT}$}&
 {\small pseudo-first order rate of adding a unit on top of a terminal
T unit}&
 {\small $\mu M^{-1}s^{-1}$}\tabularnewline
\hline
{\small $k_{gD}$}&
 {\small pseudo-first order rate of adding a unit on top of a terminal
D unit}&
 {\small $\mu M^{-1}s^{-1}$}\tabularnewline
\hline
{\small $\lambda$}&
 {\small parameter of exponential distribution (number of units)}&
 -\tabularnewline
\hline
{\small $\ell$}&
 {\small characteristic cap size (number of units)}&
 -\tabularnewline
\hline
{\small $m$}&
 {\small mean length (number of units) of MTs}&
 -\tabularnewline
\hline
{\small $n$}&
{\small coarsened step size in the cap model (number of units)}&
 -\tabularnewline
\hline
{\small $\rho$}&
 {\small $(n-1)/(\ell-1)$}&
 -\tabularnewline
\hline
{\small $K_{e}$}&
 {\small rate of the edge-induced catastrophe}&
 {\small $s^{-1}$}\tabularnewline
\hline
{\small $K_{h}$}&
 {\small rate of hydrolysis (transformation to D state) of internal
T units}&
 {\small $s^{-1}$}\tabularnewline
\hline
{\small $K_{gT}$}&
 {\small rate of adding a unit on top of a terminal T unit}&
 {\small $s^{-1}$}\tabularnewline
\hline
{\small $K_{gT}^{eff}$}&
 {\small effective rate of growing by one unit in growth phase}&
 {\small $s^{-1}$}\tabularnewline
\hline
{\small $K_{gT}^{obs}$}&
 {\small rate of growing by $n$ units in growth phase}&
 {\small $s^{-1}$}\tabularnewline
\hline
{\small $K_{gD}$}&
 {\small rate of adding a unit on top of a terminal D unit}&
 {\small $s^{-1}$}\tabularnewline
\hline
{\small $K_{n}$}&
 {\small nucleation rate of a MT}&
 {\small $s^{-1}$}\tabularnewline
\hline
{\small $K_{sT}$}&
 {\small rate of depolymerization of terminal T unit}&
 {\small $s^{-1}$}\tabularnewline
\hline
{\small $K_{sT}^{eff}$}&
 {\small effective rate of shortening by one unit in growth phase}&
 {\small $s^{-1}$}\tabularnewline
\hline
{\small $K_{sT}^{obs}$}&
 {\small rate of shortening by $n$ units in growth phase, = catastrophe
frequency}&
 {\small $s^{-1}$}\tabularnewline
\hline
{\small $K_{sD}$}&
 {\small rate of depolymerization of terminal D unit}&
 {\small $s^{-1}$}\tabularnewline
\hline
{\small $L$}&
 {\small maximal length (number of units) of MTs in domain with upper
bound}&
 -\tabularnewline
\hline
{\small $L_{x}$, $L_{y}$, $L_{z}$}&
 {\small domain sizes in the numerical simulations}&
 {\small $m$}\tabularnewline
\hline
$M_{g}(l)$&
 {\small number of MTs of length $l$ (number of units) in growth
phase}&
 -\tabularnewline
\hline
$M_{s}(l)$&
 {\small number of MTs of length $l$ (number of units) in shortening
phase}&
 -\tabularnewline
\hline
$N_{0}$&
 number of free nucleation seeds&
 -\tabularnewline
\hline
{\small $N_{MT}$}&
 {\small number of MTs, $\leq N_{n}$}&
 -\tabularnewline
\hline
{\small $N_{n}$}&
 {\small total number of nucleation seeds}&
 -  \tabularnewline
\hline
\end{tabular}\end{center}

\caption{\label{cap:Notation}Notation highlights. The dash (-) means that
the considered parameter/variable is dimensionless.}
\end{table*}
\ref{cap:Notation}. For simplicity, in our study of the role of the
boundary (e.g., cell edge) we assume that all MTs have an identical
maximal allowed length. All MTs grow from nucleation sites (there
is no spontaneous nucleation), and MTs grow at one end (usually the
so-called plus end) only. There is a fixed amount of total tubulin
in the domain. This tubulin is present in two forms: free tubulin
in the solution and polymerized tubulin constituting the MTs. Free
tubulin is taken up by growing (polymerizing) MTs and is released
back into the solution by shortening MTs. In general, free tubulin
(Tu) diffuses inside the domain. In this paper we assume that the
diffusion of free Tu is fast and does not lead to a diffusion-limited
reaction rates. This is in agreement with \cite{Odde97}. Moreover,
we assume uniform concentration of free tubulin throughout the domain
which implies instantaneous diffusion. (For the studies of the effects
of tubulin diffusion see \cite{Dogterom95,Deymier05}.)

In this paper a MT is represented at each moment in time in the form
of a 1D straight line consisting of a certain number of units of a
predefined length. Each unit belonging to a MT can be in either a
growth-prone state or a shortening-prone state. We will refer to them
as GTP (or T) state or GDP (or D) state respectively. All free tubulin
is assumed to be in a T state. When a unit joins the MT it is initially
in a T state. The probability that the internal units have hydrolyzed
(transformed to a D state) increases with time. When MT disassembles
(shortens) these D units, upon becoming terminal, have higher probability
to disassemble and return to the solution than the terminal T units.
Upon return to the solution they immediately switch to the T state.
The terminal T unit does not hydrolyze but can with a certain probability
depolymerize (drop from MT end). Incorporation of a new unit at the
MT tip triggers the hydrolysis process of the previously terminal
unit. This description seems appropriate in view of \cite{Howard03}.

The dynamics of the MTs is determined by five microscopic rates, $K_{gT}$
($K_{gD}$) and $K_{sT}$ ($K_{sD}$), which are the rates of MT growth
and shortening (i.e. adding one more unit from the solution on top
of the current terminal unit, and losing this current terminal unit
to the solution) when the terminal unit is in state T (D), and the
hydrolysis rate $K_{h}$ of the internal units which are in state
T. If the terminal unit has to hydrolyze in order to depolymerize
(and its hydrolysis rate is not faster than that for the internal
units) then $K_{sT}<K_{h}$.

For numerical simulations, shortening rates are taken to be independent
of \emph{c} while growth rates are assumed proportional to \emph{c}
at the location of MT tip: \begin{equation}
K_{gT,gD}=k_{gT,gD}c.\label{eq:Kkc}\end{equation}
 Such specific dependence of the growth and shortening rates on \emph{c},
though, is not required for many of the theoretical results we report.

When the MT reaches boundary of the domain it is not allowed to grow
any more and will eventually lose its terminal unit initiating with
certain probability a shortening phase. There are two more rates at
the domain boundary of importance in the model: rate $K_{n}$ of nucleation
from existing seeds and a rate $K_{e}$ of edge-induced catastrophe,
which can also depend on \emph{c}. Appendix \ref{app:Numerical-simulation-procedure}
contains a brief description of a numerical algorithm we used in our
simulations.

In what follows, we will impose restriction on a maximal length of
a MT (upper bound, e.g., due to a cell edge). We will call zero a
lower bound.

\subsection{Observables\label{sub:Observables}}

The standard experimental observables describing the dynamic behavior
of a single MT are derived from the MT length vs. time plot. Typical
length history plots of MTs are shown in Fig.%
\begin{figure}
\begin{center}\includegraphics[%
  width=0.39\paperwidth,
  keepaspectratio]{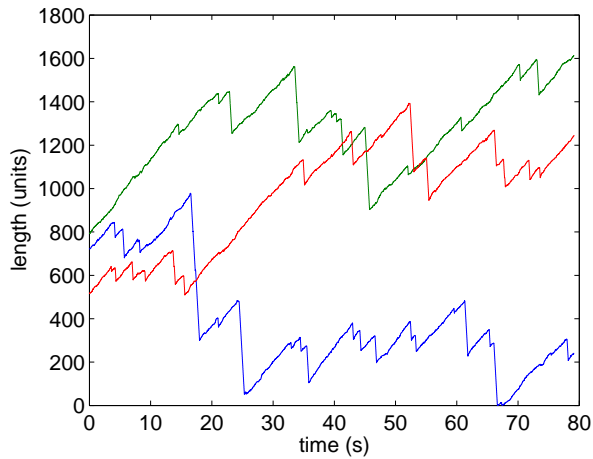}\end{center}

\caption{\label{cap:Length-history-plots}Length history plots of three arbitrarily
selected MTs. Zero time here corresponds to 80 seconds from the start
of the simulation. Here $K_{h}=10$, $c_{tot}=10$ and other parameters
are listed in Table \ref{cap:Comparison}. See also Fig. \ref{cap:Histograms-of-MT-length}.}
\end{figure}
\ref{cap:Length-history-plots}. Due to the two-state nature of the
tubulin units inside the MT, the fluctuations in this length may be
large and even in the (macroscopically) steady state each MT can repeatedly
change its length all the way from zero to some characteristic length,
or to the boundary. If no boundary is present and if free tubulin
concentration stays high enough (if $K_{gT}$ and $K_{gD}$ are high
enough) MTs can grow unbounded \cite{Dogterom93,Fygenson94,Dogterom95}.
From a sawtooth-like evolution of an MT length four parameters can
be extracted: the velocity/rate of growth, the velocity/rate of shortening,
the average time of growth before switching to shortening, and the
average time of shortening before switching to growth. The inverses
of these times define the so-called catastrophe frequency and the
rescue frequency, respectively. Note that brief growth or shortening
intervals may pass unnoticed in the analysis of experimental data.
Some models of dynamics of MTs use these four parameters as given
constants for constructing analytical solutions \cite{Dogterom93,Govindan04}
ignoring their microscopic origin. In Section \ref{sec:Cap-model}
we use microscopic rates to derive the observable growth velocity
and catastrophe frequency instead of setting them from the beginning.

\section{Cap model\label{sec:Cap-model}}

Carlier and colleagues \cite{Carlier81,Melki90,Melki96} have provided
experimental evidence that the GTP cap of the MTs is not restricted
to the units at the very tip. This suggests that the hydrolysis is
not instantaneous, a conclusion also supported by work on yeast tubulin
\cite{Davis94,Dougherty98}. To incorporate this feature into our
approach we develop a new model for the cap. In this section we show
that using this model, based on the underlying microscopic laws, one
can predict the observables: the catastrophe frequency and the velocity
(rate) in the growth phase of MTs. In other words, the microscopic
laws of the MT dynamics, governing single unit polymerization/depolymerization
and hydrolysis, can be related to macro-scale (observable) dynamics
of an MT.

When $K_{h}<\infty$, the MT cap in our model consists mainly of T
units (cf. Fig. \ref{cap:cartoon}B) and possibly of a few D units
and has some characteristic length (number of units) $\ell>1$. When
$K_{h}\rightarrow\infty$ then $\ell=1$ because only the terminal
unit is not allowed to hydrolyze and it is in a T state. Our approach
is based on coarsening the resolution in the growth phase so that
only blocks of the order of cap size, $n\sim\ell$, are added. By
catastrophe we understand the loss of the cap. In what follows we
will establish a connection between coarsened {}``observed'' rate
constants $K_{sT,gT}^{obs}$ of growing or shortening in the growth
phase, and the original rates $K_{sT,gT}$ and $K_{h}$. There is
no need to rescale $K_{sD,gD}$, if the free tubulin concentration
is not too low. Therefore, we only consider a model for the MT cap
and don't alter the rest.

Let us consider the cap model in detail. In what follows we neglect
the fluctuations in $\ell$ due to randomness in hydrolysis, and we
assume that each T unit is hydrolyzed after staying an internal unit
for a time $\Delta t_{h}=1/K_{h}$. After the rescue or nucleation
event occurs the cap begins to grow. It has time $\Delta t_{g}+\Delta t_{h}$
to elongate, where $\Delta t_{g}=1/(K_{gT}+K_{sT})$ is the time of
a single-unit step in the growth phase. After that its average length
remains constant (under assumption that no catastrophe occurs during
this time). For the catastrophe to occur the cap should be lost, due
to fluctuations in cap size and in the growth velocity \cite{Hill84a,Hill84b,Odde96}.
In our analysis we consider two scenarios for cap loss: (i) roughly
half of the cap is lost due to random nature of MT's growth, and the
second half gets hydrolyzed during this time, or (ii) the whole cap
is lost due to random fluctuations in MT growth. Keeping in mind that
the terminal unit cannot be lost as a result of hydrolysis, in description
(i) we require that $(\ell+1)/2$ and $(\ell-1)/2$ units are lost
due to fluctuations in MT growth and propagation of hydrolysis front,
respectively. We define \begin{equation}
\rho=\frac{n-1}{\ell-1}.\label{eq:rho}\end{equation}
 In case (i) $n=(\ell+1)/2$ and $\rho=1/2$, while in case (ii) $n=\ell$
and $\rho=1$. It is important to stress that $\rho$ is introduced
as a fixed parameter set \emph{a priori}, based on the scenarios of
cap loss similar to (i) and (ii). The two descriptions (i) and (ii)
determine the duration of the coarsened step \begin{equation}
\Delta t_{g}^{obs}\equiv\frac{1}{K_{gT}^{obs}+K_{sT}^{obs}}=\rho\Delta t_{h}+\Delta t_{g}\equiv\rho\cdot\frac{1}{K_{h}}+\frac{1}{K_{gT}+K_{sT}}.\label{eq:deltatg-obs}\end{equation}

For a given $n$, in order to rescale/coarsen the dynamics in the
growth phase we require that both the average velocity and the diffusion
coefficient of the MT tip remain unchanged. For a random walk on a
line, with probability \emph{p} to jump to the right and $q=1-p$
to jump to the left, the average velocity is $v=(p-q)\Delta x/\Delta t$
and the diffusion coefficient is $D=2pq\Delta x^{2}/\Delta t$. Here
$\Delta x$ is the step length and $\Delta t$ is the time per step.
In case of the original walk $p=p_{g}=K_{gT}\Delta t_{g}$, $\Delta x=1$
and $\Delta t=\Delta t_{g}$, while in case of the rescaled walk $p=K_{gT}^{obs}\Delta t_{g}^{obs}$,
$\Delta x=n$ and $\Delta t=\Delta t_{g}^{obs}$ is given by eq. (\ref{eq:deltatg-obs}).
After introducing the effective rates of adding or losing one unit
(as opposed to one block), \begin{equation}
K_{gT,sT}^{eff}\equiv nK_{gT,sT}^{obs},\label{eq:KeffKobs}\end{equation}
 and using conservation of $v$ and $D$ we obtain that \begin{equation}
K_{gT}^{eff}-K_{sT}^{eff}=K_{gT}-K_{sT}\label{eq:conserve-v}\end{equation}
\begin{equation}
\frac{K_{gT}^{eff}K_{sT}^{eff}n}{K_{gT}^{eff}+K_{sT}^{eff}}=\frac{K_{gT}K_{sT}}{K_{gT}+K_{sT}}.\label{eq:conserve-D}\end{equation}
 From eqs. (\ref{eq:deltatg-obs}) and (\ref{eq:KeffKobs}) $n/(K_{gT}^{eff}+K_{sT}^{eff})$
can be found and substituted into eq. (\ref{eq:conserve-D}), resulting
in \begin{equation}
K_{gT}^{eff}K_{sT}^{eff}=\beta,\label{eq:beta-eq}\end{equation}
 where\begin{equation}
\beta\equiv\frac{K_{gT}K_{sT}}{1+\rho\cdot{\displaystyle \frac{K_{gT}+K_{sT}}{K_{h}}}}.\label{eq:beta}\end{equation}
 After solving Eqs. (\ref{eq:conserve-v}) and (\ref{eq:beta-eq})
and choosing only positive solutions we obtain that \begin{equation}
K_{gT}^{eff}=\frac{K_{gT}-K_{sT}+\sqrt{(K_{gT}-K_{sT})^{2}+4\beta}}{2}\label{eq:KgTeff}\end{equation}
\begin{equation}
K_{sT}^{eff}=\frac{-K_{gT}+K_{sT}+\sqrt{(K_{gT}-K_{sT})^{2}+4\beta}}{2}.\label{eq:KsTeff}\end{equation}
 Expression (\ref{eq:KgTeff}) displays some expected features. Namely,
when $K_{h}\rightarrow\infty\Rightarrow K_{gT}^{eff}\rightarrow K_{gT}$
and when $K_{h}\rightarrow0\Rightarrow K_{gT}^{eff}\rightarrow(K_{gT}-K_{sT}+|K_{gT}-K_{sT}|)/2=\left[\begin{array}{cc}
K_{gT}-K_{sT}, & K_{gT}\geq K_{sT}\\
0, & K_{gT}\leq K_{sT}\end{array}\right.$. In general, $\max(0,K_{gT}-K_{sT})\leq K_{gT}^{eff}\leq K_{gT}$.

Eq. (\ref{eq:conserve-D}) combined with eqs. (\ref{eq:KgTeff}) and
(\ref{eq:KsTeff}) yields\begin{equation}
n=\frac{K_{gT}K_{sT}}{K_{gT}+K_{sT}}\frac{\sqrt{(K_{gT}-K_{sT})^{2}+4\beta}}{\beta}\label{eq:n}\end{equation}
 and eq. (\ref{eq:rho}) can now be used to determine $\ell$. On
the other hand, $\ell$ can be approximated as follows:\begin{equation}
\ell\approx K_{gT}^{eff}\Delta t_{h}+1=\frac{K_{gT}^{eff}}{K_{h}}+1,\label{eq:cap-size-estimate}\end{equation}
 where the term $K_{gT}^{eff}\Delta t_{h}$ approximates the number
of added units after the beginning of a growth phase, before the hydrolysis
front starts moving. In fact, eq. (\ref{eq:cap-size-estimate}) can
be used as a definition of $\ell$ and \emph{n} can be found from
eq. (\ref{eq:rho}). Then there is no need for eq. (\ref{eq:deltatg-obs})
as eqs. (\ref{eq:conserve-v}), (\ref{eq:conserve-D}) and (\ref{eq:cap-size-estimate})
form a closed set of equations. This approach, however, leads to a
cubic equation for $K_{gT}^{eff}$, while in the above approach we
need to solve a quadratic equation, which is much simpler. Nevertheless,
the definition of $\ell$ through eq. (\ref{eq:cap-size-estimate})
seems to work better in the limit of $K_{gT}\rightarrow0$. Namely,
substituting eq. (\ref{eq:cap-size-estimate}) into eq. (\ref{eq:rho})
yields $n=1+\rho K_{gT}^{eff}/K_{h}$ and the only non-negative solution
of this equation together with eqs. (\ref{eq:conserve-v}) and (\ref{eq:conserve-D}),
in the limit $K_{gT}\rightarrow0$, is $K_{gT}^{eff}=0$, $K_{sT}^{eff}=K_{sT}$
and $n=1$. Indeed, it seems reasonable to postulate that $n\rightarrow1$,
i.e., there is no rescaling, when $K_{gT}\rightarrow0$. This does
not follow from eqs. (\ref{eq:beta}) and (\ref{eq:n}), which lead
to $n\rightarrow1+\rho K_{sT}/K_{h}>1$ instead.

Using the above developments, it is possible to derive scaling behaviors
of various quantities as functions of, e.g., \emph{c} and $K_{h}$.
For example, substituting eqs. (\ref{eq:rho}) and (\ref{eq:cap-size-estimate})
into eq. (\ref{eq:beta}) and using eq. (\ref{eq:beta-eq}) together
with eq. (\ref{eq:KeffKobs}) leads to\begin{equation}
\begin{array}{ccc}
K_{sT}^{obs} & \approx & {\displaystyle \frac{1}{n}\cdot\frac{K_{gT}K_{sT}}{K_{gT}^{eff}+(n-1)(K_{gT}+K_{sT})}}\\
\\ & \approx & \left[\begin{array}{cc}
{\displaystyle \frac{K_{sT}}{n^{2}}}, & K_{gT}\gg K_{sT}\\
\\{\displaystyle \frac{K_{sT}}{2n(n-1)}}, & K_{gT}\gg K_{gT}^{eff}\end{array}\right.,\end{array}\label{eq:}\end{equation}
so that the catastrophe rate (frequency) $K_{sT}^{obs}$ scales as
$n^{-2}\propto\ell^{-2}$. This is characteristic of diffusive scaling
because time to catastrophe is determined by diffusive movement of
the hydrolysis front relative to the MT tip. If free tubulin concentration
\emph{c} is not too small, then eqs. (\ref{eq:n}), (\ref{eq:beta})
and (\ref{eq:Kkc}) yield $n\propto c$ and hence $K_{sT}^{obs}\propto c^{-2}$,
which is in at least qualitative agreement with previous predictions
\cite{Chen85b,Mitchison87}. The scaling $K_{sT}^{obs}\propto n^{-2}$
might have been postulated, as well, which would have led us to an
additional version of the solution of the cap model.

To provide another scaling example, let us now assume that $K_{h}$
is small and $K_{gT}\approx K_{sT}$. Then from eq. (\ref{eq:beta})
it follows that $\beta\approx K_{h}K_{sT}/(2\rho)\propto K_{h}$.
If $\beta\gg(K_{gT}-K_{sT})^{2}$, i.e., if $K_{h}$ is not too small,
then eq. (\ref{eq:n}) yields $n\propto K_{h}^{-1/2}$ and hence $\ell\propto K_{h}^{-1/2}$
as well. Notice, that it is the same scaling as derived for actin
polymers \cite[ eq.3]{Vavylonis05}.

\section{Ensemble dynamics of microtubules\label{sec:Ensemble-dynamics}}

In this section we treat cap as a single effective unit - cf. Fig.
\ref{cap:cartoon}. Thus the model essentially reduces to the two-phase
model proposed in \cite{Hill84b}. First, we rederive length distribution
of MTs, which are known in the literature. In particular, we study
the role of upper bound (e.g., cell edge). We use these results for
analyzing in the next section competition for a finite tubulin pool.
We also consider the steady state critical concentration of free Tu.

In what follows, we use either the discrete or the continuous description
of MT dynamics, whichever is convenient. We assume that the continuous
model provides a good approximation of the discrete model. The continuous
approach was discussed in \cite{Hill87,Verde92,Dogterom93} while
an analogous discrete approach was developed in \cite{Hill84b,Govindan04}.
Following \cite{Dogterom93} we write down the equations for length
distributions of MTs in growth and shortening phases in the form \begin{equation}
\partial_{t}M_{g}=-K_{sT}^{obs}M_{g}+K_{gD}M_{s}-K_{gT}^{eff}\partial_{l}M_{g}\label{eq:MgDE}\end{equation}
\begin{equation}
\partial_{t}M_{s}=K_{sT}^{obs}M_{g}-K_{gD}M_{s}+K_{sD}\partial_{l}M_{s},\label{eq:MsDE}\end{equation}
 where $M_{g,s}(l,t)$ are densities of MTs of length $l$ at time
$t$, in the growing (g) and shortening (s) phases.

Equations (\ref{eq:MgDE}) and (\ref{eq:MsDE}) can be used to describe
regular diffusion with drift, if we don't distinguish between the
phases (see also \cite{Maly02}). However, it is important to stress
that these equations don't have diffusion terms for $M_{g,s}(l,t)$
and hence switching phases back and forth is the only mechanism of
spreading of these distributions present. This is in agreement with
our simulations in the case of instantaneous hydrolysis of internal
units. Notice that diffusion terms are used in \cite{Hill84b}.

First, consider a semi-infinite domain. Equations (\ref{eq:MgDE})
and (\ref{eq:MsDE}) with the boundary condition $M_{g,s}(l=\infty)=0$
have the following steady state solution \begin{equation}
M_{g}=Ae^{-l/\lambda},\label{eq:Mg}\end{equation}
\begin{equation}
M_{s}=\frac{K_{gT}^{eff}}{K_{sD}}Ae^{-l/\lambda},\label{eq:Ms}\end{equation}
 where \begin{equation}
\frac{1}{\lambda}\equiv\frac{K_{sT}^{obs}}{K_{gT}^{eff}}-\frac{K_{gD}}{K_{sD}}.\label{eq:lambda}\end{equation}
 The necessary condition for the existence of a steady state in the
case without a boundary is given by $\lambda>0$. The prefactor $A$
is a normalization coefficient which depends on the total number of
MT nucleation seeds present and on the nucleation probability.

We now add a constraint limiting the maximal length of MTs to be $L$.
MTs cannot become longer due to a barrier, for example a cell edge,
as is often the case \emph{in vivo}, especially when the cell is in
the interphase and the MTs are relatively long. For simplicity we
assume that $L$ is identical for all MTs. We still can use eqs. (\ref{eq:MgDE})
and (\ref{eq:MsDE}) inside the domain, for $0<l<L$, and consider
a steady state. Adding up these two equations then leads to $\partial_{l}(-K_{gT}^{eff}M_{g}+K_{sD}M_{s})=0,$
which means that the spatial derivative of the flux (of the MT tips,
considered as random walkers) is zero meaning that the flux is uniform.
However, in the closed system this flux must be zero and hence\begin{equation}
M_{s}=\frac{K_{gT}^{eff}}{K_{sD}}M_{g}.\label{eq:MsMg}\end{equation}
 It follows that eqs. (\ref{eq:Mg}), (\ref{eq:Ms}) and (\ref{eq:lambda})
still hold inside the domain, except for $\lambda$ not being necessarily
positive, which is in agreement with previous work \cite{Govindan04}.
This means qualitatively that there might be a steady state distribution
of MTs in which most of them are close to the upper boundary (e.g.,
cell edge), while only a few are short.

\subsection{Critical concentration of free tubulin}

Let us consider the limiting case $1/\lambda=0$ (cf. eq. (\ref{eq:lambda})).
This defines the upper limit of the concentration of free tubulin
$c_{eq}^{\infty}$ at which the steady state in the semi-infinite
domain still exists (cf. \cite{Dogterom93}). Let us use eq. (\ref{eq:Kkc})
and define\begin{equation}
\begin{array}{ccc}
a={\displaystyle \frac{K_{sT}}{k_{gT}}}, &  & b={\displaystyle \sqrt{\frac{K_{sT}K_{sD}}{k_{gT}k_{gD}}}}.\end{array}\label{eq:ab}\end{equation}
 Because the MTs in the growth phase are less likely to shorten than
are MTs in the shortening phase, $a<b$. In general, $c_{eq}^{\infty}\in[a,b]$.
Notice that slowdown of hydrolysis reduces $c_{eq}^{\infty}(K_{h})$.
When hydrolysis is instantaneous then \emph{c} reaches its maximal
value $c_{eq}^{\infty}(\infty)=b$. When there is no hydrolysis at
all then $c_{eq}^{\infty}(K_{h})$ reaches its minimal value $c_{eq}^{\infty}(0)=a$
meaning that the average growth rate in this case, $k_{gT}c_{eq}^{\infty}(0)-K_{sT}$,
is zero.

One can get a scaling estimate of $c_{eq}^{\infty}$ if it is far
enough from both $a$ and $b$. Assume that $K_{gT}\gg K_{sT}$ and
$K_{gT}\gg K_{h}$. Using eq. (\ref{eq:Kkc}), then $K_{gT}^{eff}\approx K_{gT}\propto c$.
Now, $\rho$ is of order of 1, so that from eq. (\ref{eq:beta}) and
eq. (\ref{eq:n}) it follows that $\beta\sim K_{sT}K_{h}\ll K_{gT}^{2}$
and $n\sim K_{gT}/K_{h}\propto c/K_{h}$, respectively. Hence $K_{sT}^{obs}\approx K_{sT}/n^{2}\propto(K_{h}/c)^{2}$.
Substituting these scaling relations into $1/\lambda=0$ and using
eq. (\ref{eq:lambda}) yields $c_{eq}^{\infty}\propto\sqrt{K_{h}}$.

Notice that if there are no rescues, $k_{gD}=0$, then $c_{eq}^{\infty}$
is infinite (see also \cite{Mitchison87}) and unbounded growth can
not happen. This is so because without rescues the MT depolymerizes
completely after the catastrophe, no matter how long it was before.

\section{Competition for tubulin\label{sec:Competition-for-tubulin}}

In Section \ref{sec:Ensemble-dynamics} we have shown the existence
of a steady state distribution of MTs inside a domain. It is conceivable
that by sensing and controlling free tubulin concentration and the
number of MTs the cell regulates MT dynamics, as suggested in \cite{Mitchison87,Gregoretti}.
In what follows we show in detail how to determine a steady state
concentration of free tubulin \emph{c}, which is the key to finding
steady state characteristics of MTs in a closed system. The main goal
of this section is to derive expressions for the average number of
units per MT $m$, and a number of MTs $N_{MT}$ as functions of $c$
and the other parameters. These functions are needed to determine
\emph{c} from the conservation of total tubulin. Because \emph{m}
can not extend beyond the domains' boundary, and because $N_{MT}$
is less or equal than a number of nucleation sites $N_{n}$, the number
of polymerized units in a bounded domain stays restricted as the amount
of total Tu grows.

In what follows we assume that the total amount of tubulin is constant
and we consider bounded domain of volume $V$. We also assume instantaneous
diffusion so that the concentration of free tubulin is uniform throughout
the domain. Hence,

\begin{equation}
N_{tot}=N_{free}+mN_{MT},\label{eq:}\end{equation}
 where $N_{tot}$ is a total number of tubulin units in the domain
and $N_{free}$ is a number of free tubulin units. By dividing this
formula by $V$ we obtain expression for concentrations measured in
micromolars ($\mu M$) \begin{equation}
c_{tot}=c+\frac{mN_{MT}}{10^{-3}N_{A}V},\label{eq:ctot}\end{equation}
 where $V$ is given in $m^{3}$ and $1\mu M$ is equal to $10^{-6}N_{A}$
units per liter or $10^{-3}N_{A}$ units per $m^{3}$; $N_{A}\approx6.022\times10^{23}\textrm{mol}^{-1}$
is the Avogadro's constant. We will use this expression for studying
MTs in cases of unbounded and bounded domains.

\subsection{Unbounded domain}

It has been shown in Section \ref{sec:Ensemble-dynamics} that the
steady state distribution of MT lengths in this case is exponential
as described by eqs. (\ref{eq:Mg}) and (\ref{eq:Ms}) with $\lambda$
representing mean length of a MT \begin{equation}
m=\lambda.\label{eq:meqlambda}\end{equation}
 To find $N_{MT}$ for a given $N_{n}$, we use a balance equation
for the number of available nucleation sites $N_{0}\equiv N_{n}-N_{MT}$:\begin{equation}
K_{sT}^{obs}M_{g}(l=1)+K_{sD}M_{s}(l=1)=K_{n}N_{0},\label{eq:nucleation-balance}\end{equation}
 where $K_{n}$ is a nucleation rate, which in general depends on
$c$. The left-hand side of eq. (\ref{eq:nucleation-balance}) describes
the rate of production of available nucleation sites by completely
depolymerizing MTs. The first term represents those MTs which experience
a catastrophe, while the second term stands for those MTs which are
already in the shortening phase. Using (\ref{eq:Mg}), (\ref{eq:Ms}),
and assuming that $\lambda\gg1$, yields\begin{equation}
A(K_{sT}^{obs}+K_{gT}^{eff})\approx K_{n}N_{0}.\label{eq:}\end{equation}
 In addition, approximating summation by integration as follows\begin{equation}
N_{MT}=\sum_{l=1}^{\infty}\left(M_{g}(l)+M_{s}(l)\right)\approx\left(1+\frac{K_{gT}^{eff}}{K_{sD}}\right)A\lambda\label{eq:}\end{equation}
 results in \begin{equation}
N_{MT}\approx\frac{N_{n}}{1+{\displaystyle \frac{K_{sT}^{obs}+K_{gT}^{eff}}{\lambda K_{n}(1+K_{gT}^{eff}/K_{sD})}}}.\label{eq:NMT}\end{equation}
 Substituting eqs. (\ref{eq:meqlambda}), (\ref{eq:NMT}) and (\ref{eq:lambda})
into eq. (\ref{eq:ctot}), and using dependence of the rates on concentration
\emph{c}, eq. (\ref{eq:Kkc}), yields an equation $c_{tot}=F(c)$,
at steady state. This equation relates free tubulin concentration
\emph{c} and total concentration $c_{tot}$, as a function of all
given parameters.

\subsection{Bounded domain}

Here again we impose limitation on the maximal possible length of
MTs not to exceed \emph{L}. In the case of a bounded domain a steady
state solution of eqs. (\ref{eq:MgDE}) and (\ref{eq:MsDE}) can be
calculated, and then $m$ and $N_{MT}$ are determined in a way similar
to the previous case. Notice that the steady state does exist even
if $c>c_{eq}^{\infty}$. Since $0\leq m\leq L$ and $0\leq N_{MT}\leq N_{n}$
always hold, the second term in eq. (\ref{eq:ctot}) is non-negative
and bounded. Therefore, when \emph{c} goes from 0 to $\infty$ so
does $c_{tot}$. If the right-hand side of eq. (\ref{eq:ctot}) monotonically
increases with \emph{c} then there is a unique physically meaningful
\emph{c} for each $c_{tot}$, at steady state.

It is shown in Appendix \ref{app:Competition-with-edge} that now\begin{widetext}\begin{equation}
N_{MT}\approx\frac{N_{n}}{1+{\displaystyle \frac{K_{sT}^{obs}+K_{gT}^{eff}}{K_{n}\left[(1+K_{gT}^{eff}/K_{sD})(1-e^{-L/\lambda})\lambda+(K_{gT}^{eff}/K_{e})(1+K_{gD}/K_{sD})e^{-L/\lambda}\right]}}}\label{eq:NMT-edge}\end{equation}
 and\begin{equation}
m\approx\frac{(1+K_{gT}^{eff}/K_{sD})\lambda(\lambda-e^{-L/\lambda}(L+\lambda))+L(K_{gT}^{eff}/K_{e})(1+K_{gD}/K_{sD})e^{-L/\lambda}}{(1+K_{gT}^{eff}/K_{sD})\lambda(1-e^{-L/\lambda})+(K_{gT}^{eff}/K_{e})(1+K_{gD}/K_{sD})e^{-L/\lambda}}.\label{eq:m}\end{equation}
 \end{widetext}When $\lambda>0$ and $L\rightarrow\infty$ eqs. (\ref{eq:NMT-edge})
and (\ref{eq:m}) reduce to eqs. (\ref{eq:NMT}) and (\ref{eq:meqlambda}),
respectively. When $K_{gT}^{eff}\rightarrow0$ and $K_{gD}\rightarrow0$,
then $\lambda\downarrow0$ (so that $\lambda>0$), $N_{MT}\rightarrow0$
and $m\rightarrow0$. When $K_{gT}^{eff}\rightarrow\infty$ and $K_{gD}\rightarrow\infty$,
then $\lambda\uparrow0$ (so that $\lambda<0$), $N_{MT}\rightarrow N_{n}$
and $m\rightarrow L$, as expected.

\section{Discussion of the results\label{sec:Discussion}}

\subsection{Comparison with existing cap models\label{sub:Comparison-of-cap-models}}

The cap model presented in this paper differs from the approaches
used in \cite{Hill84a,Hill84b,Flyvbjerg94,Flyvbjerg96}. First, we
don't postulate catastrophe frequency and growth velocity, or derive
them from numerical simulations, as was done in \cite{Hill84a,Hill84b}.
Instead, we analytically derive these macroscopic rates from small-scale
rates (such as chemical rate constants). Second, we employ only spontaneous
hydrolysis and don't use induced, or vectorial, hydrolysis; both types
of hydrolysis were used in \cite{Flyvbjerg94,Flyvbjerg96}. Our model
agrees with the experimental data analyzed in \cite{Flyvbjerg94,Flyvbjerg96}
as can be seen from the main panel in Fig.%
\begin{figure}
\begin{center}\includegraphics[%
  width=0.4\paperwidth,
  keepaspectratio]{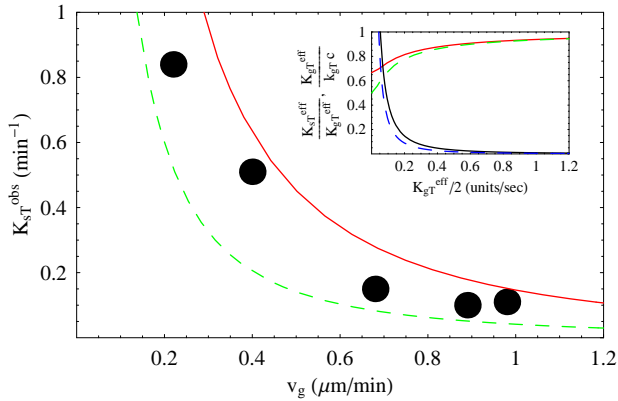}\end{center}

\caption{\label{cap:Fcat-experim} Frequency of catastrophe $K_{sT}^{obs}$
as a function of MT growth velocity $v_{g}$. Dots represent experimental
data \cite{Drechsel92,Flyvbjerg94}. Solid and dashed lines correspond
to $\rho=1/2$ and $\rho=1$, respectively, and were obtained using
the model (Section \ref{sec:Cap-model}). Inset: Two upper curves
on the right are $K_{gT}^{eff}/K_{gT}$ and two lower curves on the
right are $K_{sT}^{eff}/K_{gT}^{eff}$.}
\end{figure}
\ref{cap:Fcat-experim}. Specifically, the predicted dependence of
the catastrophe frequency on MT growth velocity is in agreement with
experimental data.

The cap dynamics in the model proposed by Flyvbjerg \emph{et al.}
\cite{Flyvbjerg94,Flyvbjerg96} is modeled by addition of tubulin
from the solution to the MT tip. This addition (polymerization) is
faster than the propagation of the induced hydrolysis front (low end
of the cap). Therefore, the cap length would grow infinite were it
not for the spontaneous hydrolysis at some point inside the cap. When
it occurs, the cap is redefined as an interval between this spontaneous
hydrolysis point and the MT tip. In this way, the average cap size
can be kept constant at steady state. According to Flyvbjerg \emph{et
al.} \cite{Flyvbjerg94,Flyvbjerg96}, catastrophe occurs when this
cap is lost, and it is postulated that the remaining GTP-Tu units
below the cap are not capable of rescuing the MT. This assumption
is made in order to allow for catastrophe to occur. Otherwise, in
many cases the rescue would immediately follow the cap loss. While
this picture is widely accepted, in our alternative model the picture
is even simpler. We use only one type of hydrolysis and we don't need
to make any additional assumptions. In our model, there is a hydrolysis
front due to spontaneous hydrolysis of old enough units (see Section
\ref{sec:Cap-model}). The velocity of this front is governed by the
\emph{age} of the units inside the MT and hence it is always approximately
equal (with fluctuations) to the growth velocity. Faster growth velocity
leads to a longer cap, reducing the catastrophe frequency.

Dilution experiments have shown that sharp reduction in the concentration
of free Tu to low or zero values results in collapse of the MTs after
a certain delay. Importantly, this delay is practically independent
of the initial free Tu concentration \cite{Walker91}. Flyvbjerg \emph{et
al.} explain this phenomenon by arguing that the dilution results
in domination of spontaneous hydrolysis which regulates the waiting
time before the collapse. Therefore, this time is almost independent
of the initial cap size. Our model yields the following simple explanation.
When concentration of free Tu becomes very low, two events must occur
for the cap to disappear. The terminal unit should be lost (rate $K_{sT}$)
and the next unit should hydrolyze (rate $K_{h}$). If this next to
last unit is old enough to hydrolyze then, with high probability,
the rest of the cap has already hydrolyzed.

Dilution experiments reported in \cite{Walker91} and cited in \cite{Flyvbjerg94}
determine the average waiting time before the catastrophe as roughly
5-10s. Therefore, in Fig. \ref{cap:Fcat-experim} we set $K_{h}=K_{sT}=0.15s^{-1}$.
If the shortening velocity ($K_{sD}$) is much larger than $K_{sT}$,
and if the loss of the terminal unit is conceptualized as a two-stage
process of hydrolysis and then falling, the equality $K_{h}=K_{sT}$
would indicate that the hydrolysis rate of the terminal unit equals
the hydrolysis rate of the internal unit. Notice that we were not
able to fit the catastrophe frequency data if $K_{sT}$ and $K_{h}$
were significantly different. In the dilution experiments spatial
resolution was about $0.25\mu m$ \cite{Walker91}, which is about
30 heterodimers (per protofilament), so that actual waiting time before
losing the terminal unit (heterodimer) might be faster than the reported
waiting time before the collapse. We used eq. (\ref{eq:Kkc}) with
$k_{gT}=0.3\mu M^{-1}s^{-1}$, however any value of $k_{gT}$ can
be used, because it enters the formulas only through $K_{gT}=k_{gT}c$
and there is no explicit \emph{c}-dependence in the figure. The unit
length is taken to be the length of one heterodimer of Tu, 8nm and
hence $1\, unit/s\approx0.5\,\mu m/min$. Therefore $v_{g}\,(\mu m/min)\approx K_{gT}^{eff}/2\,(units/s)$.
In the inset of Fig. \ref{cap:Fcat-experim} we show the values of
$K_{sT}^{eff}/K_{gT}^{eff}$ and $K_{gT}^{eff}/K_{gT}$. It is seen
that for $v_{g}>0.2$ one has $K_{sT}^{obs}<K_{sT}^{eff}\ll K_{gT}^{eff}$
and hence $v_{g}\propto K_{gT}^{eff}\propto c$.

\subsection{Competition for tubulin and the edge effect\label{sub:Competition-finite-Kh}}

Here we study competition for a limited pool of free tubulin and combine
it together with the cap model. At steady state, the dependence of
free tubulin concentration, \emph{c}, on the total tubulin concentration,
$c_{tot}$, is governed by tubulin mass conservation eq. (\ref{eq:ctot}).
The resulting value of \emph{c}, in turn, defines the dynamics of
MTs and their ensemble distributions.

In the case of an unbounded spatial domain, we have reproduced the
prediction of the Oosawa and Kasai model \cite{Oosawa62} including
existence of a critical concentration of free tubulin $c_{eq}^{\infty}$
(see thin lines in Fig.%
\begin{figure}
\begin{center}\includegraphics[%
  width=0.4\paperwidth,
  keepaspectratio]{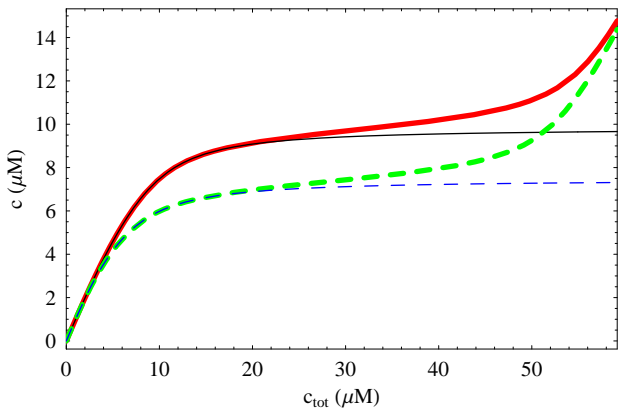}\end{center}

\caption{\label{cap:c-vs-ctot} Steady state concentration of free vs. total
Tu in the unbounded (thin lines) and bounded (thick lines) domains.
For a bounded domain maximal MT length \emph{L} is roughly 1400 units.
For a bounded domain \emph{c} reaches its asymptotic value $c_{eq}^{\infty}(K_{h})$.
Hydrolysis rate is $K_{h}=10s^{-1}$. Full and dashed lines correspond
to $\rho=1/2$ and $\rho=1$, respectively. Other parameters are $N_{n}=200$,
$V=10^{-17}m^{3}=10^{-8}\mu\textrm{L}$, $k_{gT}=5\mu M^{-1}s^{-1}$,
$K_{sT}=5s^{-1}$, $k_{gD}=0.5\mu M^{-1}s^{-1}$, $K_{sD}=500s^{-1}$,
$K_{n}=K_{gD}$, $K_{e}=K_{sT}$, and we use eq. (\ref{eq:Kkc}).}
\end{figure}
\ref{cap:c-vs-ctot}). The steady state cannot exist above this value.
Depending on the choice of parameters, the transition from almost
linear growth of \emph{c,} $c\approx c_{tot}$ for low $c_{tot}$,
to the asymptotic value $c=c_{eq}^{\infty}$ can be made sharp or
smooth. (In Oosawa model, this transition is assumed to be sharp meaning
that when $c_{tot}<c_{eq}^{\infty}$ there are no MTs and when $c_{tot}>c_{eq}^{\infty}$
all the excess tubulin above $c_{eq}^{\infty}$ goes into polymerized
state.) When the probability of rescue is zero then $c_{eq}^{\infty}\rightarrow\infty$,
our model describes the situation considered in \cite{Mitchison87}.

If there is an upper bound on MT lengths, then at sufficiently high
concentrations of total tubulin, this bound inhibits further polymerization
of MTs (edge effect). Hence, the steady state free tubulin concentration
can rise above its critical value. This edge effect is demonstrated
by thick lines in Fig. \ref{cap:c-vs-ctot} and was first discussed
in \cite{Gregoretti}. Under our reference conditions, it is seen
that for $c_{tot}>20\mu M$ the edge starts playing an important role\textbf{.}
At sufficiently high $c_{tot}$, the edge reestablishes linear growth
of \emph{c} with respect to $c_{tot}$. The implications of this effect
are as follows. If $c_{tot}$ is high enough, the MTs grow persistently
up until hitting the edge which triggers their catastrophe. This is
consistent with recent experimental observations of persistent growth
of MTs \emph{in vivo} \cite{Komarova02}. From our model it follows
that by changing the number of nucleation sites while keeping the
total amount of Tu constant the cell could regulate the transition
between mitotic (short) and interphase (long) MTs.

Next we compare in Table%
\begin{table*}
\begin{center}\begin{tabular}{|c|c|c|c|c|c|c|c|c|c|c|c|c|}
\hline 
\multicolumn{3}{|c|}{Parameters}&
\multicolumn{4}{c|}{Simulated results}&
\multicolumn{6}{c|}{Theoretical estimates: $n=(\ell+1)/2$, $n=\ell$ (see text)}\tabularnewline
\hline
$K_{h}$&
 $c_{tot}$&
 $N_{n}$&
 $c$&
 $m$&
 $\frac{\# T}{N_{n}}$&
 $N_{MT}$&
 $c_{eq}^{\infty}$&
 $c$&
 $\lambda$&
 $\ell$&
 $\frac{K_{gT}^{eff}}{K_{h}}+1$&
 $N_{MT}$\tabularnewline
\hline
{\scriptsize $\infty$}&
 {\scriptsize 36}&
 {\scriptsize 2000}&
 {\scriptsize 31.32}&
 {\scriptsize 1416}&
 {\scriptsize 0.759}&
 {\scriptsize 1990}&
 {\scriptsize 31.62}&
 {\scriptsize 31.28}&
 {\scriptsize 1430}&
 {\scriptsize 1}&
 {\scriptsize 1}&
 {\scriptsize 1989}\tabularnewline
\hline
{\scriptsize $\infty$}&
 {\scriptsize 28}&
 {\scriptsize 3000}&
 {\scriptsize 27.46}&
 {\scriptsize 117}&
 {\scriptsize 0.734}&
 {\scriptsize 2814}&
 {\scriptsize -''-}&
 {\scriptsize 27.48}&
 {\scriptsize 112}&
 {\scriptsize 1}&
 {\scriptsize 1}&
 {\scriptsize 2797}\tabularnewline
\hline
{\scriptsize $\infty$}&
 {\scriptsize 28}&
 {\scriptsize $10^{4}$, T seed}&
 {\scriptsize 26.48}&
 {\scriptsize 92.7}&
 {\scriptsize 0.780}&
 {\scriptsize 9908}&
 {\scriptsize -''-}&
 {\scriptsize 26.53}&
 {\scriptsize 89.5}&
 {\scriptsize 1}&
 {\scriptsize 1}&
 {\scriptsize 9909}\tabularnewline
\hline
{\scriptsize $10^{5}$}&
 {\scriptsize 36}&
 {\scriptsize 2000}&
 {\scriptsize 31.36}&
 {\scriptsize 1401}&
 {\scriptsize 0.997}&
 {\scriptsize 1995}&
 {\scriptsize 31.60, 31.57}&
 {\scriptsize 31.25, 31.23}&
 {\scriptsize 1437, 1444}&
 {\scriptsize 1.0015}&
 {\scriptsize 1.0016}&
 {\scriptsize 1989}\tabularnewline
\hline
{\scriptsize 100}&
 {\scriptsize 30}&
 {\scriptsize 2000}&
 {\scriptsize 20.78}&
 {\scriptsize 2788}&
 {\scriptsize 1.75}&
 {\scriptsize 1992}&
 {\scriptsize 21.05, 17.38}&
 {\scriptsize 20.92, 17.29}&
 {\scriptsize 2742, 3836}&
 {\scriptsize 2.00, 1.82}&
 {\scriptsize 2.03, 1.84}&
 {\scriptsize 1994, 1996}\tabularnewline
\hline
{\scriptsize 30}&
 {\scriptsize 15}&
 {\scriptsize 2000}&
 {\scriptsize 12.7}&
 {\scriptsize 694}&
 {\scriptsize 2.69}&
 {\scriptsize 1969}&
 {\scriptsize 14.7, 11.5}&
 {\scriptsize 13.9, 11.2}&
 {\scriptsize 347, 1139}&
 {\scriptsize 3.16, 2.72}&
 {\scriptsize 3.22, 2.77}&
 {\scriptsize 1952, 1985}\tabularnewline
\hline
{\scriptsize 10}&
 {\scriptsize 10}&
 {\scriptsize 2000}&
 {\scriptsize 7.51}&
 {\scriptsize 758}&
 {\scriptsize 3.94}&
 {\scriptsize 1974}&
 {\scriptsize 9.85, 7.48}&
 {\scriptsize 9.02, 7.18}&
 {\scriptsize 305, 857}&
 {\scriptsize 5.04, 4.12}&
 {\scriptsize 5.16, 4.20}&
 {\scriptsize 1945, 1980}\tabularnewline
\hline
{\scriptsize 10}&
 {\scriptsize 5}&
 {\scriptsize 2000}&
 {\scriptsize 4.91}&
 {\scriptsize 35.1}&
 {\scriptsize 2.25}&
 {\scriptsize 1538}&
 {\scriptsize -''-}&
 {\scriptsize 4.95, 4.85}&
 {\scriptsize 21.9, 52.1}&
 {\scriptsize 3.06, 2.98}&
 {\scriptsize 3.20, 3.07}&
 {\scriptsize 1420, 1722}\tabularnewline
\hline
{\scriptsize 3}&
 {\scriptsize 5}&
 {\scriptsize 2000}&
 {\scriptsize 3.94}&
 {\scriptsize 327}&
 {\scriptsize 5.66}&
 {\scriptsize 1954}&
 {\scriptsize 6.08, 4.59}&
 {\scriptsize 4.64, 3.99}&
 {\scriptsize 115, 312}&
 {\scriptsize 7.22, 6.09}&
 {\scriptsize 7.42, 6.21}&
 {\scriptsize 1868, 1952}\tabularnewline
\hline
{\scriptsize 1}&
 {\scriptsize 4}&
 {\scriptsize 2000}&
 {\scriptsize 2.56}&
 {\scriptsize 443}&
 {\scriptsize 8.51}&
 {\scriptsize 1958}&
 {\scriptsize 3.90, 2.99}&
 {\scriptsize 3.24, 2.63}&
 {\scriptsize 237, 420}&
 {\scriptsize 12.5, 9.42}&
 {\scriptsize 12.77, 9.53}&
 {\scriptsize 1941, 1970}\tabularnewline
\hline
{\scriptsize 0.3}&
 {\scriptsize 5}&
 {\scriptsize 2000}&
 {\scriptsize 1.795}&
 {\scriptsize 971}&
 {\scriptsize 14.0}&
 {\scriptsize 1988}&
 {\scriptsize 2.50, 2.00}&
 {\scriptsize 2.31, 1.86}&
 {\scriptsize 817, 950}&
 {\scriptsize 23.6, 16.1}&
 {\scriptsize 23.7, 16.1}&
 {\scriptsize 1986, 1990}\tabularnewline
\hline
{\scriptsize 0.1}&
 {\scriptsize 3}&
 {\scriptsize 2000}&
 {\scriptsize 1.379}&
 {\scriptsize 496}&
 {\scriptsize 19.8}&
 {\scriptsize 1970}&
 {\scriptsize 1.782, 1.510}&
 {\scriptsize 1.573, 1.368}&
 {\scriptsize 433, 494}&
 {\scriptsize 32.1, 21.5}&
 {\scriptsize 31.6, 20.8}&
 {\scriptsize 1982, 1988}\tabularnewline
\hline
{\scriptsize 0.03}&
 {\scriptsize 3}&
 {\scriptsize 2000}&
 {\scriptsize 1.192}&
 {\scriptsize 553}&
 {\scriptsize 32.9}&
 {\scriptsize 1969}&
 {\scriptsize 1.368, 1.234}&
 {\scriptsize 1.262, 1.163}&
 {\scriptsize 526, 555}&
 {\scriptsize 50.7, 33.2}&
 {\scriptsize 48.5, 31.1}&
 {\scriptsize 1991, 1994}\tabularnewline
\hline
{\scriptsize 0.01}&
 {\scriptsize 2}&
 {\scriptsize 2000}&
 {\scriptsize 1.080}&
 {\scriptsize 286}&
 {\scriptsize 42.5}&
 {\scriptsize 1937}&
 {\scriptsize 1.179, 1.112}&
 {\scriptsize 1.097, 1.058}&
 {\scriptsize 273, 284}&
 {\scriptsize 65.6, 43.2}&
 {\scriptsize 58.4, 37.0}&
 {\scriptsize 1992, 1995 }\tabularnewline
\hline
\end{tabular}\end{center}

\caption{\label{cap:Comparison} Comparison of simulated and theoretical results.
All values reported here are at or very close to steady state. Cells
containing two numbers show theoretical predictions for $\rho=1/2$
and 1, respectively. Domain size is $L_{x}=L_{y}=10^{-4}$ and $L_{z}=10^{-7}$
(m). It is ensured that the domain is long enough so that the MTs
don't reach the boundary for the given parameters. Here $k_{gT}=5$,
$K_{sT}=5$, $k_{gD}=0.5$, $K_{sD}=500$, $K_{gT,gD}=k_{gT,gD}c$.
For nucleation rate we use $K_{n}=K_{gD}$, i.e., a D seed, except
for the third line, where it is a T seed, $K_{n}=K_{gT}$. $\# T/N_{n}$
is an estimate for the cap size - see Section \ref{sub:Competition-finite-Kh}.
Initially all Tu is free and its concentration is $c_{tot}$. Notice
that $c<c_{eq}^{\infty}$.}
\end{table*}
\ref{cap:Comparison} Monte Carlo simulations (see Appendix \ref{app:Numerical-simulation-procedure})
with the results obtained by using our continuous model. We choose
large domain size so that MTs never reach the boundary and $m=\lambda$
(cf. Eq. (\ref{eq:meqlambda})). We run simulations for different
parameter sets until steady state is reached and then determine free
tubulin concentration, number of MTs and their mean length, and estimate
the cap size. Recall that only MTs in growth phase have caps. Therefore,
we estimate the cap size {\small $\ell$} of MTs in our simulations
by {\small $\ell\sim(1+K_{gT}^{eff}/K_{sD})\cdot\# T/N_{MT}$,} where
$\# T$ is the number of polymerized Tu units in T state. We also
use eq. (\ref{eq:MsMg}). For the parameter values used in Table \ref{cap:Comparison}
simplified formula for the estimated cap size $\ell\sim\# T/N_{n}$
has been used.

Table \ref{cap:Comparison} also contains theoretical estimates corresponding
to the simulated values discussed above. In addition, the table includes
theoretical estimates of $c_{eq}^{\infty}$ and of the cap size based
on eq. (\ref{eq:cap-size-estimate}). In most cases the simulated
results lie in between our two theoretical approximations, for $\rho=1/2$
and $\rho=1$ respectively, in agreement with the model description
of cap evolution. These approximations are given as two adjacent numbers
in the cells of the table displaying our theoretical estimates. When,
however, $K_{h}$ becomes small, and \emph{c} approaches \emph{a}
(eq. (\ref{eq:ab})), our approximations seem to consistently overestimate
the number of MTs, $N_{MT}$. This should be improved by rescaling
the rest of the rates, $K_{gD}$, $K_{sD}$, $K_{n}$ and $K_{e}$,
which is outside of the scope of this paper.

\subsection{Non-steady-state phenomena}

Our numerical simulations are also capable of reproducing and explaining
two other phenomena observed in experiments. Specifically, (i) in
some cases the non-exponential MT length distribution and (ii) decaying
oscillations in free tubulin concentration have been observed.

\emph{MT length distributions.} It is often mentioned in the literature
that the steady state length distribution of MTs observed in the experiments
is either exponential or bell-shaped \cite{Gliksman93,Gliksman92,Cassimeris86,Verde90,Bolterauer99a,Bolterauer99b}.
Exponential distribution agrees with our model. Inability to obtain
a bell-shaped distribution seems to indicate a limitation to our model.
While we don't exclude the possibility that some rates might depend
on the MT length, as proposed by \cite{Bolterauer99a,Bolterauer99b},
or on time spent in a given phase \cite{Odde95,Odde98}, we suggest
two alternative ways of obtaining bell-shaped distributions under
certain conditions. First, one should be careful in determining when
the system reaches the steady state in the experiment or simulation.
As our simulations demonstrate, the system reaches the constant free
tubulin concentration and MTs reach constant mean length relatively
quickly. (The number of MTs does not change much henceforth.) By that
time the MT length histogram is often bell-shaped as illustrated in
Fig.%
\begin{figure*}
\begin{center}\begin{tabular}{cc}
\includegraphics[%
  width=0.39\paperwidth,
  keepaspectratio]{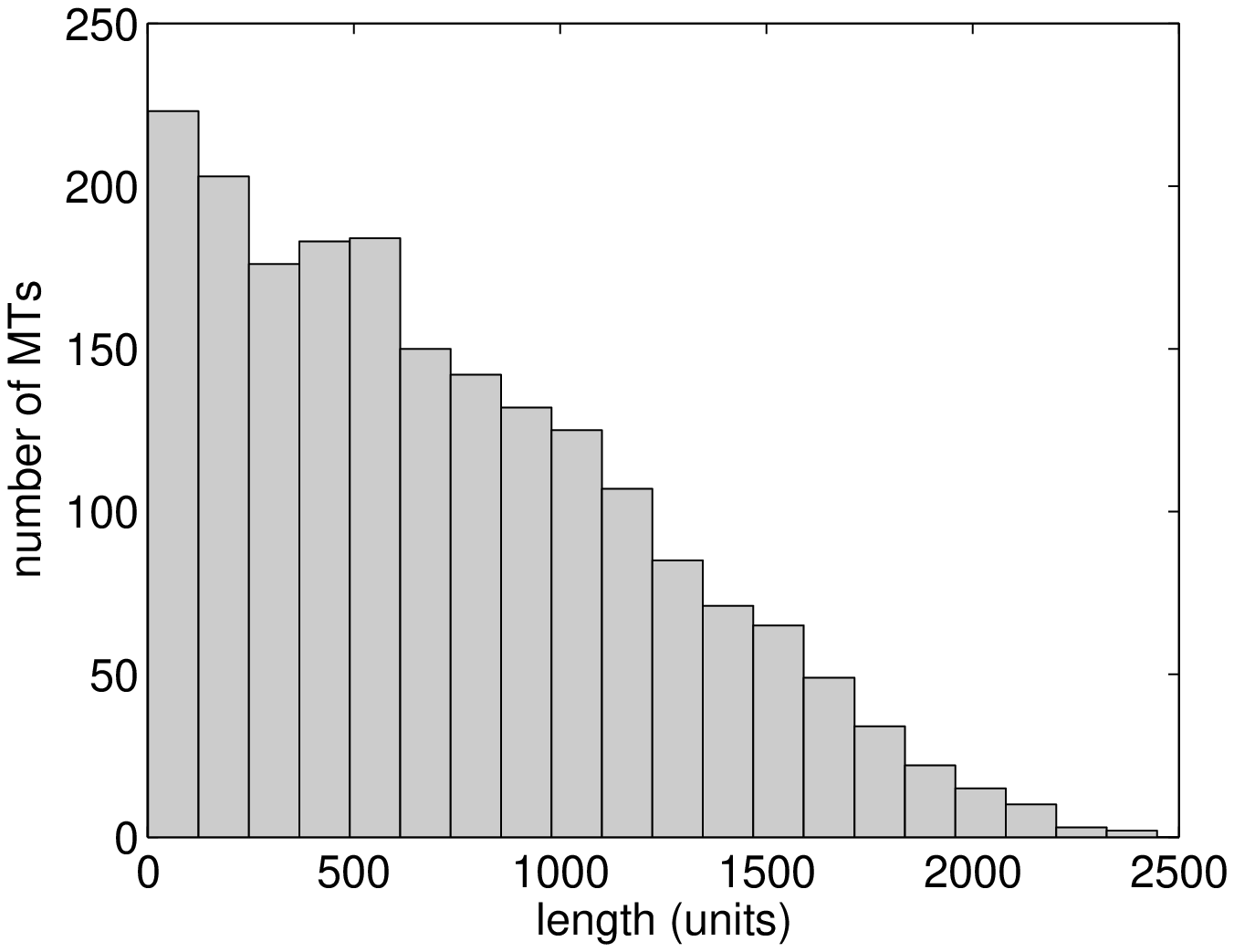}&
\includegraphics[%
  width=0.39\paperwidth,
  keepaspectratio]{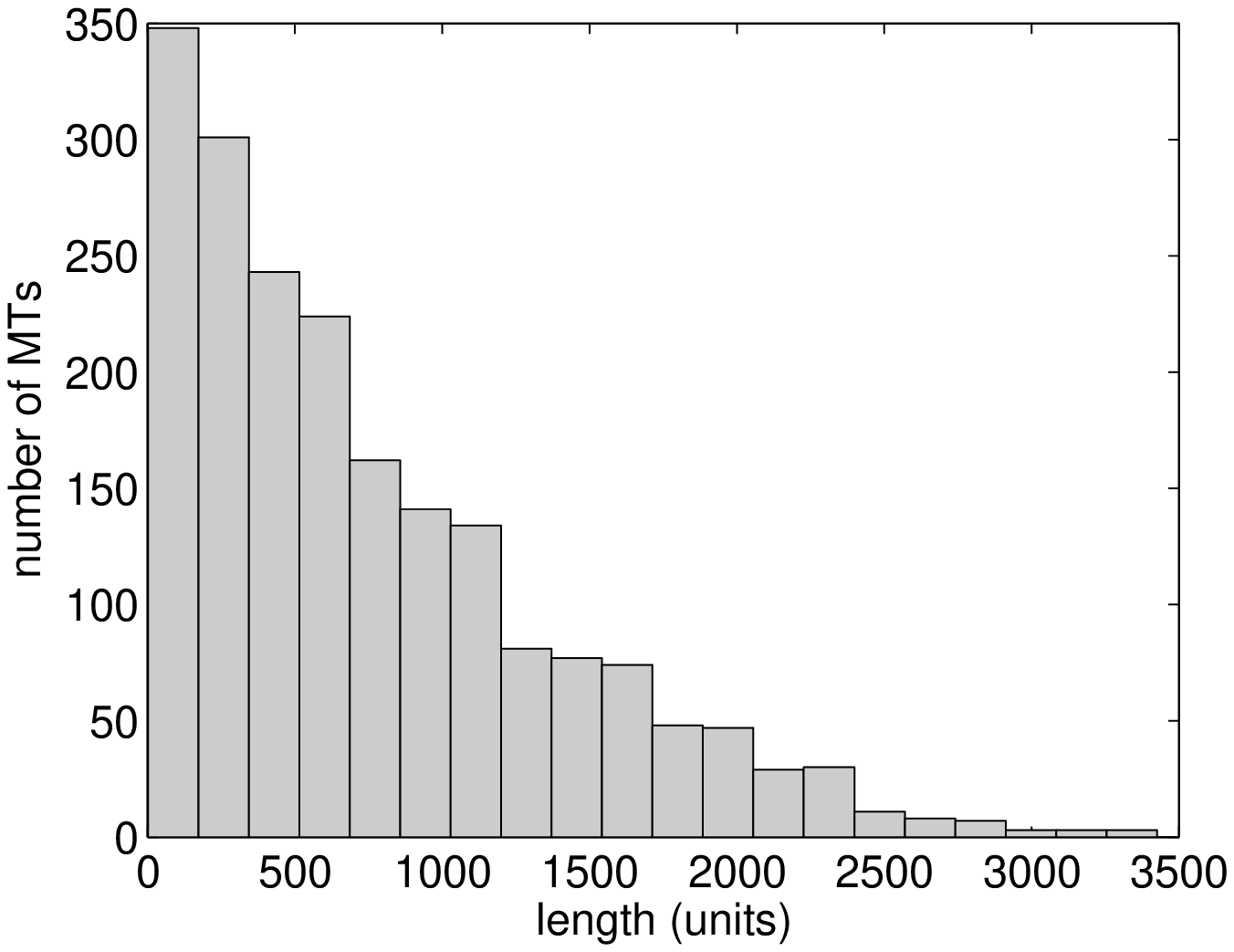}\tabularnewline
\end{tabular}\end{center}

\caption{\label{cap:Histograms-of-MT-length}Histograms of MT lengths after
80 (left) and 160 (right) seconds from the beginning of polymerization.
The concentration of free Tu in both cases is close to steady state
value: 7.63 and 7.51 $\mu$M, respectively. Although the concentration
has reached steady state after 80 sec, the length distribution of
MTs is still changing. Here $K_{h}=10s^{-1}$, $c_{tot}=10\mu M$
and rest of the parameters are specified in Table \ref{cap:Comparison}.}
\end{figure*}
\ref{cap:Histograms-of-MT-length} and Fig.%
\begin{figure}
\begin{center}\includegraphics[%
  width=0.39\paperwidth,
  keepaspectratio]{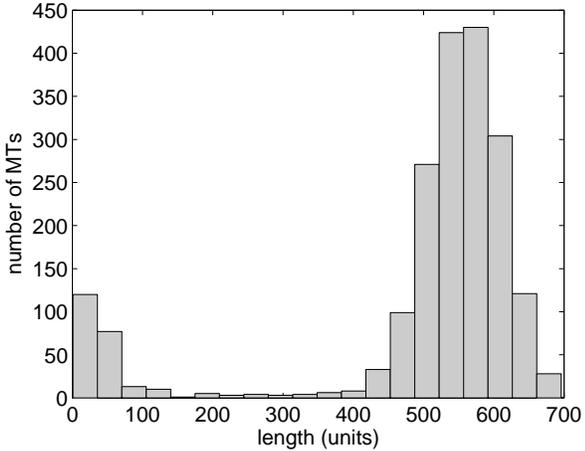}\end{center}

\caption{\label{cap:Histogram-of-MT-length2} Histogram of MT lengths after
150 seconds from the beginning of polymerization. The concentration
of free Tu is 1.382 $\mu M$ and has reached its steady state value
(to within random fluctuations). $K_{h}=0.1s^{-1}$, $c_{tot}=3\mu M$
and rest of the parameters are specified in Table \ref{cap:Comparison}.
Note the bimodal character of the distribution.}
\end{figure}
\ref{cap:Histogram-of-MT-length2}. This can be explained as follows.
When MTs start growing from nucleation sites there is an excess of
free tubulin. Therefore, the growth is originally unbounded leading
to a Gaussian shape. If the cell edge (upper boundary) is far away,
in the course of this growth the free tubulin concentration drops
and reaches its steady state value. At this time the shape can still
be close to a Gaussian (cf. \cite[ fig. 4]{Gliksman93}). This is
followed by a long process of a shape change of the MT length histogram
with free tubulin being constant. Eventually, this results in an exponential
shape and in system reaching true steady state. This situation is
well described in \cite{OShaughnessy03a,OShaughnessy03b}.

A bell-shaped distribution can be also obtained in a bounded domain
when the steady state concentration of free tubulin is high enough
for unbounded growth of MTs if it were not for a cell edge (i.e.,
if $c>c_{eq}^{\infty}$). In this situation we predict positive exponential
distribution of MT lengths, in the case when all MTs can reach \emph{identical}
maximal length restricted by the edge, consistent with \cite{Govindan04,Maly02}.
As can be seen in experiments, however, MTs are curved and cell shape
is not ideally spherical or circular, so that different MTs experience
different restrictions (e.g., \cite{Cassimeris86,Komarova02}). This
can lead to an MT length histogram of a bell-shaped form, in true
steady state.

\emph{Oscillations.} In some of the simulations we observed an overshoot
in free Tu concentration before the steady state was reached. Moreover,
in some cases there were \emph{slight} oscillations of free Tu concentration,
as shown in Fig.%
\begin{figure}
\begin{center}\includegraphics[%
  width=0.39\paperwidth,
  keepaspectratio]{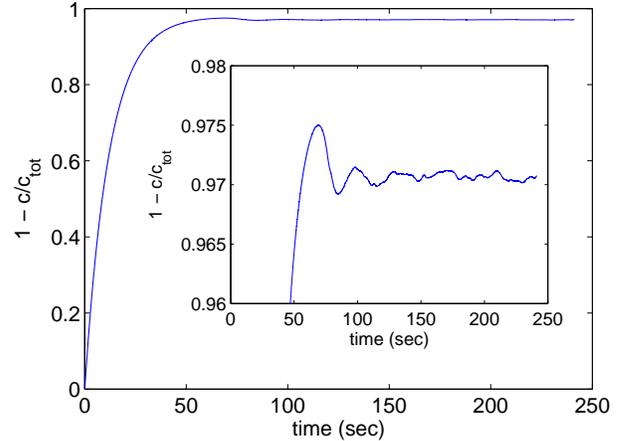}\end{center}

\caption{\label{cap:Small-oscillations} Small oscillations in the amount
of polymerized Tu observed in numerical simulations for $K_{h}=0.1s^{-1}$,
$c_{tot}=50\mu M$ and $N_{n}=10^{4}$. Other parameters are as in
Table \ref{cap:Comparison}. Inset is a blowup of the region of oscillations.}
\end{figure}
\ref{cap:Small-oscillations}. Overshoots and \emph{large} oscillations
have been reported and modeled in the literature \cite{Carlier87,Chen87,Bayley89,Jobs97,Bolterauer99b,Sept99,Sept00,Deymier05}.
It is believed that slow conversion of D- into T-tubulin in solution,
after the depolymerization, is the key to understanding of such oscillations.
These oscillations occur if initial free Tu concentration is sufficiently
large.

Free D-tubulin cannot polymerize. In our model we assume its conversion
into T-tubulin to be instantaneous. We also use linear (and not higher
order) dependence of nucleation rate on free Tu concentration, and
a fixed number of nucleation sites. It is remarkable that under these
restrictive assumptions the model produced some oscillations. We suggest
the following explanation for their existence. If the hydrolysis is
slow enough, the MTs grow quickly in the beginning, resulting in a
large cap. When free Tu concentration changes quickly, the cap needs
a relatively long time to adjust. This leads to a delayed response,
and possibly to oscillations. Hence it might be that the ability to
produce oscillations is inherent to MT structure, and that it can
be magnified under certain experimental conditions.

\section{Conclusions\label{sec:Conclusions}}

In this paper we analyze a novel model of MT dynamics in a domain
bounded by the cell edge which involved competition of individual
MTs for tubulin. The model is based on a linear 1D approximation of
a MT structure. We consider the role of the boundary and extended
the model to incorporate finite hydrolysis.

One of the main results of the paper is an establishment of a link,
by using analytical formulas, between microscopic parameters describing
polymerization/depolymerization and hydrolysis of individual units,
and macroscopic (observable) characteristics of the MT dynamics and
ensemble. We demonstrate how to approximate macroscopic steady state
behavior of MTs using microscopic rates and vice versa, extract microscopic
rates from macroscopic behavior. Hence, it is possible to analytically
and quantitatively predict the effect of changes in microscopic parameters
on observable features, as well as to deduce microscopic changes from
observed changes in macroscopic behavior, when relevant geometry and
chemistry is taken into account.

We consider the cell edge, and derive expressions characterizing MT
distributions in the case of a finite pool of tubulin units. We perform
Monte Carlo simulations to compare with our theoretical results, and
find a good agreement between the two.

The key ingredient in establishing a link between micro- and macro-parameters
is the cap model, which allows one to replace the actual cap consisting
of many units by an effective single unit. The cap model behavior
agrees with experiments measuring catastrophe frequency as a function
of free tubulin concentration as well as with dilution experiments.

We also use computational Monte Carlo model to provide an explanation
for a non-exponential MT length distributions observed in experiments.

We plan to develop in future a detailed model for an individual MT
which will take into account the MT protofilament structure, and to
use this model for investigating effects of various MAPs on the dynamic
instability.

~

\begin{acknowledgments}
This research was supported in part by the NIH grant 1 RO1 GM065420:
Supplement for the Study of Complex Biological Systems.
\end{acknowledgments}
\appendix

\section{Computational model \label{app:Numerical-simulation-procedure}}

In what follows we provide a short description of the numerical algorithm
used for our simulations, which slightly differs from that used in
\cite{Gregoretti}. At time zero, MTs begin to grow from the nucleation
seeds. At each simulation step, the time of this step is calculated
by defining the rates of change in length (either growth or shortening)
for each MT. If the MT tip is in T (D) state then this rate of change
is $K=K_{gT(gD)}+K_{sT(sD)}$. We demand that the maximal average
number of changes for each MT will be 1, which means that we find
a maximal value among all $K$ and then set $\Delta t=1/\max_{MTs}\{ K\}$
{[}a technical note - we set $\Delta t$ slightly below this value
because the rate $K$ for a given MT can change slightly as \emph{c}
is affected after each MT changes its length; these changes in \emph{c}
are usually very small{]}. Then, in general, each MT will have a chance
to grow, shorten, or retain its length during $\Delta t$. We don't
allow distribution of possible number of length changes for an MT
during $\Delta t$ (only zero or one change is possible). After the
length of each MT is updated, we update \emph{c} accordingly; after
updating the lengths of all MTs, the hydrolysis cycle runs through
all internal units of all MTs. The probability that a unit will hydrolyze
during time $\Delta t$ is taken as $1-e^{-K_{h}\Delta t}$, assuming
Poisson statistics.

All MTs have their first unit in D state and this unit cannot be lost
- this constitutes a simple nucleation seed with lower growth probability
than when the MT tip is in T state (these units are not counted when
calculating the lengths of the MTs). Similarly, when the edge is relevant
we can assume $K_{e}=K_{sT}$. Such choices are made purely to reduce
the number of parameters in the system, and are not essential for
our purposes.

\section{Explicit solution in the bounded domain \label{app:Competition-with-edge}}

In what follows we obtain expressions for $N_{MT}$ and $m$ in the
bounded domain at the steady state. Notice that in our model all the
MTs of the maximum length \emph{L} are technically in the growing
phase, because their terminal unit can never become internal and therefore
does not hydrolyze. (These MTs cannot grow because of the edge.) The
edge-induced catastrophe rate $K_{e}$ will be governed by the smallest
of the rates $K_{sT}$, $K_{h}$. The discrete version of eqs. (\ref{eq:MgDE})
and (\ref{eq:MsDE}) determining a steady state at the boundary \begin{equation}
0=-K_{e}M_{g}(L)+K_{gD}M_{s}(L-1)+K_{gT}^{eff}M_{g}(L-1),\label{eq:MgDE-edge}\end{equation}
\begin{equation}
0=K_{e}M_{g}(L)-K_{gD}M_{s}(L-1)-K_{sD}M_{s}(L-1).\label{eq:MsDE-edge}\end{equation}
 After summing up these two equations we recover \begin{equation}
M_{s}(L-1)=(K_{gT}^{eff}/K_{sD})M_{g}(L-1),\label{eq:MsMg-edge}\end{equation}
 which is already known (cf. (\ref{eq:MsMg})) and so one of these
equations is superfluous. Another way to find $M_{g,s}$ is to write
the general solution of (\ref{eq:MgDE}), (\ref{eq:MsDE}): $M_{g}=Ae^{-z/\lambda}+B$,
$M_{s}=(K_{gT}^{eff}/K_{sD})Ae^{-z/\lambda}+(K_{sT}^{obs}/K_{gD})B$
and plug it in (\ref{eq:MsMg-edge}) yielding $B=0$ unless $(K_{sT}^{obs}K_{sD})/(K_{gT}^{eff}K_{gD})=1$
in which case $B$ is arbitrary. But this last condition implies $\lambda\rightarrow\infty$
and hence $M_{g,s}$ are constant inside the domain. At the lower
domain boundary, eq. (\ref{eq:nucleation-balance}) still holds.

The number of MTs is given now by \begin{equation}
N_{MT}=\sum_{l=1}^{L-1}(M_{g}(l)+M_{s}(l))+M_{g}(L)=N_{n}-N_{0}.\label{eq:}\end{equation}
 Using eqs. (\ref{eq:Mg}), (\ref{eq:Ms}), (\ref{eq:nucleation-balance})
and (\ref{eq:MgDE-edge}) and replacing the summation by the integration
from 0 to \emph{L} we can determine \emph{A} and hence $N_{MT}$,
which is given in eq. (\ref{eq:NMT-edge}).

Similarly, mean MT length is \begin{equation}
m=\frac{{\displaystyle \sum_{l=1}^{L-1}}l(M_{g}(l)+M_{s}(l))+LM_{g}(L)}{{\displaystyle \sum_{l=1}^{L-1}}(M_{g}(l)+M_{s}(l))+M_{g}(L)}\label{eq:}\end{equation}
 leading to eq. (\ref{eq:m}).

%\bibliographystyle{unsrt}
%\bibliography{Model}

\begin{thebibliography}{10}

\bibitem{Mitchison84}
Tim Mitchison and Marc Kirschner.
\newblock Dynamic instability of microtubule growth.
\newblock {\em Nature}, 312(15 Nov.):237--242, 1984.

\bibitem{Hill84a}
Terrell~L. Hill and Yi~der Chen.
\newblock Phase changes at the end of a microtubule with a gtp cap.
\newblock {\em Proc. Natl. Acad. Sci. USA}, 81:5772--5776, 1984.

\bibitem{Howard03}
Joe Howard and Anthony~A. Hyman.
\newblock Dynamics and mechanics of the microtubule plus end.
\newblock {\em Nature}, 422(17 Apr.):753--758, 2003.

\bibitem{Carlier81}
Marie-France Carlier and Dominique Pantaloni.
\newblock Kinetic analysis of guanosine 5'-triphosphate hydrolysis associated
  with tubulin polymerization.
\newblock {\em Biochemistry}, 20:1918--1924, 1981.

\bibitem{Hill84b}
Terrell~L. Hill.
\newblock Introductory analysis of the gtp-cap phase-change kinetics at the end
  of a microtubule.
\newblock {\em Proc. Natl. Acad. Sci. USA}, 81:6728--6732, 1984.

\bibitem{Chen85a}
Yi~der Chen and Terrell~L. Hill.
\newblock Monte {C}arlo study of the gtp cap in a five-start helix model of a
  microtubule.
\newblock {\em Proc. Natl. Acad. Sci. USA}, 82:1131--1135, 1985.

\bibitem{Dogterom93}
Marileen Dogterom and Stanislas Leibler.
\newblock Physical aspects of the growth and regulation of microtubule
  structures.
\newblock {\em Phys. Rev. Lett.}, 70(9):1347--1350, 1993.

\bibitem{Bolterauer99a}
H.~Bolterauer, H.-J. Limbach, and J.~A. Tuszy\'nski.
\newblock Models of assembly and disassembly of individual microtubules:
  stochastic and averaged equations.
\newblock {\em J. Biological Physics}, 25:1--22, 1999.

\bibitem{Bolterauer99b}
H.~Bolterauer, H.-J. Limbach, and J.~A. Tuszy\'nski.
\newblock Microtubules: strange polymers inside the cell.
\newblock {\em Bioelectrochemistry and Bioenergetics}, 48:285--295, 1999.

\bibitem{Gliksman93}
N.~R. Gliksman, R.~V. Skibbens, and E.~D. Salmon.
\newblock How the transition frequencies of microtubule dynamic instability
  (nucleation, catastrophe, and rescue) regulate microtubule dynamics in the
  interphase mitosis: analysis using a {Monte Carlo} computer simulation.
\newblock {\em Molecular Biology of the Cell}, 4:1035--1050, 1993.

\bibitem{Govindan04}
Bindu~S. Govindan and William~B. {Spillman, Jr.}
\newblock Steady states of microtubule assembly in a confined geometry.
\newblock {\em Phys. Rev. E}, 70:032901, 2004.

\bibitem{Bayley89}
Peter~M. Bayley, Maria~J. Schilistra, and Stephen~R. Martin.
\newblock A simple formulation of microtubule dynamics: quantitative
  implications of the dynamic instability of microtubule populations in vivo
  and in vitro.
\newblock {\em J. Cell Sci.}, 93:241--254, 1989.

\bibitem{Chen85b}
Yi~der Chen and Terrell~L. Hill.
\newblock Theoretical treatment of microtubules disappearing in solution.
\newblock {\em Proc. Natl. Acad. Sci. USA}, 82:4127--4131, 1985.

\bibitem{Freed02}
Karl~F. Freed.
\newblock Analytical solution for steady-state populations in the self-assembly
  of microtubules from nucleating sites.
\newblock {\em Phys. Rev. E}, 66:061916, 2002.

\bibitem{Maly02}
Ivan~V. Maly.
\newblock Diffusion approximation of the stochastic process of microtubule
  assembly.
\newblock {\em Bull. Math. Biol.}, 64:213--238, 2002.

\bibitem{Flyvbjerg94}
Henrik Flyvbjerg, Timothy~E. Holy, and Stanislas Leibler.
\newblock Stochastic dynamics of microtubules: a model for caps and
  catastrophes.
\newblock {\em Phys. Rev. Lett.}, 73(17):2372--2375, 1994.

\bibitem{Flyvbjerg96}
Henrik Flyvbjerg, Timothy~E. Holy, and Stanislas Leibler.
\newblock Microtubule dynamics: caps, catastrophes and coupled hydrolysis.
\newblock {\em Phys. Rev. E}, 54(5):5538--5560, 1996.

\bibitem{Janosi02}
Imre~M. J\'anosi, Denis Chr\'etien, and Henrik Flyvbjerg.
\newblock Structural microtubule cap: stability, catastrophe, rescue and third
  state.
\newblock {\em Biophysical J.}, 83:1317--1330, 2002.

\bibitem{VanBuren02}
Vincent VanBuren, David~J. Odde, and Lynne Cassimeris.
\newblock Estimates of lateral and longitudinal bond energies within the
  microtubule lattice.
\newblock {\em Proc. Natl. Acad. Sci. USA}, 99(9):6035--6040, 2002; correction:
  $ibid.$ 101(41), 14989, 2004.

\bibitem{VanBuren05}
Vincent VanBuren, Lynne Cassimeris, and David~J. Odde.
\newblock Mechanochemical model of microtubule structure and self-assembly
  kinetics.
\newblock {\em Biophysical J.}, 89(5):2911--2926, 2005.

\bibitem{Stukalin04}
Evgeny~B. Stukalin and Anatoly~B. Kolomeisky.
\newblock Simple growth models of rigid multifilament biopolymers.
\newblock {\em J. Chem. Phys.}, 121(2):1097--1104, 2004.

\bibitem{Molodtsov05}
Maxim~I. Molodtsov, Elena~A. Ermakova, Emmanuil~E. Shnol, Ekaterina~L.
  Grishchuk, J.~Richard McIntosh, and Fazly~I. Ataullakhanov.
\newblock A molecular-mechanical model of the microtubule.
\newblock {\em Biophysical J.}, 88:3167--3179, 2005.

\bibitem{Gregoretti}
Ivan~V. Gregoretti, Gennady Margolin, Mark~S. Alber, and Holly~V. Goodson.
\newblock Microtubule dynamics: Monte carlo model predicts emergent properties.
\newblock (in preparation).

\bibitem{Odde97}
David~J. Odde.
\newblock Estimation of the diffusion-limited rate of microtubule assembly.
\newblock {\em Biophysical J.}, 73:88--96, 1997.

\bibitem{Dogterom95}
M.~Dogterom, A.~C. Maggs, and S.~Leibler.
\newblock Diffusion and formation of microtubule asters: physical processes
  versus biochemical regulation.
\newblock {\em Proc. Natl. Acad. Sci. USA}, 92:6683--6688, 1995.

\bibitem{Deymier05}
P.~A. Deymier, Y.~Yang, and J.~Hoying.
\newblock Effect of tubulin diffusion on polymerization of microtubules.
\newblock {\em Phys. Rev. E}, 72:021906, 2005.

\bibitem{Fygenson94}
Deborah~Kuchnir Fygenson, Erez Braun, and Albert Libchaber.
\newblock Phase diagram of microtubules.
\newblock {\em Phys. Rev. E}, 50(2):1579--1588, 1994.

\bibitem{Melki90}
Ronald Melki, Marie-France Carlier, and Dominique Pantaloni.
\newblock Direct evidence for \protect{GTP and GDP-P\_i} intermediates in
  microtubule assembly.
\newblock {\em Biochemistry}, 29:8921--8932, 1990.

\bibitem{Melki96}
Ronald Melki, St\'ephane Fievez, and Marie-France Carlier.
\newblock Continuous monitoring of \protect{P\_i} release following nucleotide
  hydrolysis in actin or tubulin assembly using
  2-amino-6-mercapto-7-methylpurine ribonucleoside and purine-nucleoside
  phosphorylase as an enzyme-linked assay.
\newblock {\em Biochemistry}, 35:12038--12045, 1996.

\bibitem{Davis94}
Ashley Davis, Carleton~R. Sage, Cynthia~A. Dougherty, and Kevin~W. Farrell.
\newblock Microtubule dynamics modulated by guanosine triphosphate hydrolysis
  activity of $\beta$-tubulin.
\newblock {\em Science}, 264(5160):839--842, 1994.

\bibitem{Dougherty98}
Cynthia~A. Dougherty, Richard~H. Himes, Leslie Wilson, and Kevin~W. Farrell.
\newblock Detection of \protect{GTP and P\_i in wild-type and mutated yeast
  microtubules: implications for the role of the GTP/GDP-P\_i} cap in
  microtubule dynamics.
\newblock {\em Biochemistry}, 37(31):10861--10865, 1998.

\bibitem{Odde96}
David~J. Odde, Helen~M. Buettner, and Lynne Cassimeris.
\newblock Spectral analysis of microtubule assembly dynamics.
\newblock {\em AIChE J.}, 42(5):1434--1442, 1996.

\bibitem{Mitchison87}
T.~J. Mitchison and M.~W. Kirschner.
\newblock Some thoughts on the partitioning of tubulin between monomer and
  polymer under conditions of dynamic instability.
\newblock {\em Cell Biophysics}, 11:35--55, 1987.

\bibitem{Vavylonis05}
Dimitrios Vavylonis, Qingbo Yang, and Ben O'Shaughnessy.
\newblock Actin polymerization kinetics, cap structure, and fluctuations.
\newblock {\em Proc. Natl. Acad. Sci. USA}, 102(24):8543--8548, 2005.

\bibitem{Hill87}
T.~L. Hill.
\newblock {\em Linear aggregation theory in cell biology}.
\newblock New York, Springer-Verlag, 1987.

\bibitem{Verde92}
Fulvia Verde, Marileen Dogterom, Ernst Stelzer, Eric Karsenti, and Stanislas
  Leibler.
\newblock Control of microtubule dynamics and length by cyclin a- and cyclin
  b-dependent kinases in {X}enopus egg extracts.
\newblock {\em J. Cell Biol.}, 118(5):1097--1108, 1992.

\bibitem{Drechsel92}
D.~N. Drechsel, A.~A. Hyman, M.~H. Cobb, and M.~W. Kirschner.
\newblock Modulation of the dynamic instability of tubulin assembly by the
  microtubule-associated protein tau.
\newblock {\em Mol. Biol. Cell}, 3:1141--1154, 1992.

\bibitem{Walker91}
R.~A. Walker, N.~K. Pryer, and E.~D. Salmon.
\newblock Dilution of individual microtubules observed in real time in vitro:
  evidence that cap size is small and independent of elongation rate.
\newblock {\em J. Cell Biol.}, 114(1):73--81, 1991.

\bibitem{Oosawa62}
F.~Oosawa and M.~Kasai.
\newblock A theory of linear and helical aggregations of macromolecules.
\newblock {\em J. Mol. Biol.}, 4(Jan):10--21, 1962.

\bibitem{Komarova02}
Yulia~A. Komarova, Ivan~A. Vorobjev, and Gary~G. Borisy.
\newblock Life cycle of mts: persistent growth in the cell interior, asymmetric
  transition frequencies and effects of the cell boundary.
\newblock {\em J. Cell Science}, 115:3527--3539, 2002.

\bibitem{Gliksman92}
Neal~R. Gliksman, Stephen~F. Parsons, and E.~D. Salmon.
\newblock Okadaic acid induces interphase to mitotic-like microtubule dynamic
  instability by inactivating rescue.
\newblock {\em J. Cell Biol.}, 119(5):1271--1276, 1992.

\bibitem{Cassimeris86}
Lynne~U. Cassimeris, Patricia Wadsworth, and E.~D. Salmon.
\newblock Dynamics of microtubule depolymerization in monocytes.
\newblock {\em J. Cell Biol.}, 102:2023--2032, 1986.

\bibitem{Verde90}
Fulvia Verde, {Jean-claude} Labb\'e, Marcel Dor\'ee, and Eric Karsenti.
\newblock Regulation of microtubule dynamics by cdc2 protein kinase in
  cell-free extracts of xenopus eggs.
\newblock {\em Nature}, 343:233--238, 1990.

\bibitem{Odde95}
David~J. Odde, Lynne Cassimeris, and Helen~M. Buettner.
\newblock Kinetics of microtubule catastrophe assessed by probabilistic
  analysis.
\newblock {\em Biophysical J.}, 69:796--802, 1995.

\bibitem{Odde98}
David~J. Odde and Helen~M. Buettner.
\newblock Autocorrelation function and power spectrum of two-state random
  processes used in neurite giudance.
\newblock {\em Biophysical J.}, 75:1189--1196, 1998.

\bibitem{OShaughnessy03a}
Ben O'Shaughnessy and Dimitrios Vavylonis.
\newblock The ultrasensitivity of living polymers.
\newblock {\em Phys. Rev. Lett.}, 90(11):118301, 2003.

\bibitem{OShaughnessy03b}
B.~O'Shaughnessy and D.~Vavylonis.
\newblock Dynamics of living polymers.
\newblock {\em Eur. Phys. J. E}, 12:481--496, 2003.

\bibitem{Carlier87}
M.~F. Carlier, R.~Melki, D.~Pantaloni, T.~L. Hill, and Y.~Chen.
\newblock Synchronous oscillations in microtubule polymerization.
\newblock {\em Proc. Natl. Acad. Sci. USA}, 84:5257--5261, 1987.

\bibitem{Chen87}
Yi~der Chen and Terrell~L. Hill.
\newblock Theoretical studies on oscillations in microtubule polymerization.
\newblock {\em Proc. Natl. Acad. Sci. USA}, 84:8419--8423, 1987.

\bibitem{Jobs97}
Elmar Jobs, Dietrich~E. Wolf, and Henrik Flyvbjerg.
\newblock Modeling microtubule oscillations.
\newblock {\em Phys. Rev. Lett.}, 79(3):519--522, 1997.

\bibitem{Sept99}
D.~Sept.
\newblock Model for spatial microtubule oscillations.
\newblock {\em Phys. Rev. E}, 60(1):838--841, 1999.

\bibitem{Sept00}
D.~Sept and J.~A. Tuszy\'nski.
\newblock A {Landau-Ginzburg} model of the co-existence of free tubulin and
  assembled microtubules in nucleation and oscillations phenomena.
\newblock {\em J. Biological Physics}, 26:5--15, 2000.

\end{thebibliography}

\end{document}